\documentclass[12pt,preprint]{aastex}

\usepackage{graphicx}
\usepackage{amsmath}
\usepackage{amssymb}

\title{The distances of short-hard GRBs and the SGR connection}
\author{Ehud Nakar$^1$, Avishay Gal-Yam$^{1,2}$, Tsvi Piran$^{1,3}$ and Derek B. Fox$^1$}
\affil{\begin{flushleft}1. Division of Physics, Mathematics and
Astronomy, Caltech,
Pasadena, 91125, USA \\
2. Hubble fellow \\
3. Racah Institute for Physics, The Hebrew University, Jerusalem,
91904, ISRAEL\end{flushleft} }
\begin{document}
%\maketitle

\begin{abstract}
We present a search for nearby ($D \lesssim 100$ Mpc) galaxies in
the error boxes of six well-localized short-hard gamma-ray bursts
(GRBs). None of the six error boxes reveals the presence of a
plausible nearby host galaxy. This allows us to set lower limits on
the distances and, hence, the isotropic-equivalent energy of these
GRBs. Our lower limits are around $1 \times 10^{49}$ erg (at
$2\sigma$ confidence level); as a consequence, some of the
short-hard GRBs we examine would have been detected by BATSE out to
distances greater than 1~Gpc and therefore constitute a bona fide
cosmological population. Our search is partially motivated by the
December 27, 2004 hypergiant flare from SGR~1806$-$20, and the
intriguing possibility that short-hard GRBs are extragalactic events
of a similar nature. Such events would be detectable with BATSE to a
distance of 50~Mpc, and their detection rate should be comparable to
the actual BATSE detection rate of short-hard GRBs. The failure of
our search, by contrast, suggests that such flares constitute less
than 15\% of the short-hard GRBs ($<40\%$ at $95\%$-confidence). We
discuss possible resolutions of this discrepancy.

\end{abstract}

%%%%%%%%%%%%%%%%%%%%%%%%%%%%%%%%%%%%%%%%%%%%%%%%%%

\section{Introduction}

After a decade of rapid progress in our understanding of gamma-ray
bursts (GRBs; see Piran 2004 for a recent review), the origin of
short-hard GRBs is still a mystery. The short-hard GRB class
(Kouveliotou et al.\ 1993) makes up a quarter of the entire GRB
population observed by BATSE
\footnote{http://www.batse.msfc.nasa.gov/batse/} with an all-sky
detection rate of $\approx 170 \;\rm y^{-1}$ (Meegan et al., 1997).
The revolution in our understanding of the long-duration GRBs,
triggered by the detection of their long-lived counterpart
(afterglow) emission, has so far passed over the short-hard GRBs,
lacking similar afterglow detections\footnote{
The likely X-ray afterglow of the short-hard burst GRB 050509b was
discovered by {\it SWIFT} (Gehrels et al. 2005) while this paper was
being refereed. We briefly discuss the implications of this event
in $\S~5$.}. With no afterglow detections, we have only indirect clues
about the typical distance to short-hard GRBs: the almost-isotropic
sky distribution (Briggs et al.\
1996; Balazs et al.1998; Magliocchetti et al.\ 2003) and small value
of $<V/V_{max}>$ (Katz \& Canel 1996; Schmidt 2001; Guetta \& Piran
2004) for the population suggests a cosmological origin. Even so,
the distance and energy scale of the events remains highly
uncertain.

Soft gamma-ray repeaters (SGRs) are compact sources of persistent
X-ray emission and repeating bursts of soft gamma-rays (see Woods \&
Thompson 2004 for a review). They are believed to be
highly-magnetized young neutron stars (known as magnetars; Duncan \&
Thompson 1992; Paczy\'nski 1992; Katz 1994; Thompson \& Duncan 1995,
see also Katz 1982 for related early ideas). At infrequent
intervals, these sources emit powerful flares of high energy
radiation, much larger than their usual bursts, releasing more than
$10^{44}$ ergs in the form of gamma-rays alone. These ``giant
flares'' are characterized by a short ($\lesssim 0.5$ sec) and very
intense burst that is followed by a long ($\sim 100$ sec) pulsating
soft tail. During $\sim 30$ years of observations, two of the four
confirmed SGRs (SGR~0526$-$66 and SGR~1900+14) produced a single
giant flare each.

On December 27, 2004 a third source, SGR~1806$-$20, erupted in a
giant flare (Borkowski et al.\ 2004; Mazets et al.\ 2005; Hurley et
al.\ 2005; Palmer et al.\ 2005). The initial pulse fluence was
about $1 \;\rm erg/cm^2$ . At a distance of 15~kpc (Corbel \&
Eikenberry 2004; see however Figer et al. 2004; Cameron et al. 2005;
$\S~5$) this fluence corresponds to a total energy output of $3
\times 10^{46}$ erg, about a hundred times the energy of the two
previous giant flares, making this a hypergiant flare (hereafter
HGF). As seen in previous giant flares this event showed a short and
intense pulse followed by a long pulsating soft tail.

Duncan (2001) pointed out that the initial hard pulse of an
extragalactic giant flare would be detected by BATSE as a short-hard
GRB.  He considered flares with a luminosity similar to that of the
giant flare of SGR~0526$-$66, finding that they would be detected to a
distance of $\sim$10~Mpc and could account for only a small fraction
of the short GRBs observed by BATSE. Eichler (2002) suggested that
``total field reversal'' in a magnetar could produce an HGF with an
energy output of $\sim 10^{46}$ to $10^{47}$ erg in a catastrophic
event involving the near-total destruction of the magnetic
dipole moment of the star. Such an HGF would be detected as a short-hard GRB out
to a distance of $\sim 100$ Mpc. However, according to Eichler's
model, an HGF consumes most of the exterior magnetic energy of the
magnetar and can occur only once or twice during the SGR's lifetime,
implying an HGF rate of $10^{-3} \;\rm y^{-1}$ per Milky Way-size
Galaxy. Eichler (2002) therefore also concluded that HGFs would
constitute only a small fraction of the short GRBs in the BATSE
sample.

The observation of an HGF with an energy larger than $10^{46}$ erg
after only $\sim 30$ years of observations in our Galaxy has forced
a revision of this picture. Boggs et al.\ (2005) reiterated the
possible connection between HGFs and short GRBs. Dar (2005) pointed
out that an HGF rate of $\sim 3 \times 10^{-2}\; \rm y^{-1}$ in the
Milky Way implies  an observed all-sky rate of roughly one short GRB
per day at BATSE sensitivity. This rate is comparable to the
observed rate of short GRBs, suggesting that all short GRBs may be
HGFs from SGRs.

In this letter we report on a search for nearby star-forming host
galaxies in the error boxes of six well-localized short-hard GRBs.
We use our limits on the brightness of any such host galaxies to
derive minimal distances and isotropic-equivalent energy outputs for
these bursts. We use the measured flux of the Dec. 27 HGF in the
BATSE detection window (50--300 keV), verified by our own
independent estimation derived using the observed backscatter of HGF
X-rays from the moon (Golenetskii et al.\ 2004), to estimate the
BATSE detection rate of extragalactic HGFs and to compare this
prediction with the results of our search.

In the sections below, we estimate the energy output of the recent
HGF from SGR 1806-20 in the
BATSE detection window (\S 2), calculate the expected BATSE detection
rate of extragalactic HGFs (\S 3), and describe our search for nearby
host galaxies of short-hard GRBs (\S 4). We conclude with a discussion
of the implications of our work for theories of the short-hard GRBs
and, in particular, their relation to the SGRs (\S 5).  A Hubble
constant of $H_{\circ}=70$ km/sec/Mpc is assumed throughout the paper.

%%%%%%%%%%%%%%%%%%%%%%%%%%%%%%%%%%%%%%%%%%%%%%%%%%

\section{The fluence of the SGR~1806$-$20 hypergiant flare}\label{sec
  SGR}

The Coranas-F spacecraft
\footnote{http://coronas.izmiran.rssi.ru/F/}, a solar activity
observatory in a low-earth orbit, was occulted by the Earth at the
time of the flare. The moon, at this time, was full with respect to
Coranas-F and the flare. This coincidence enabled Helicon, a
$\gamma$-ray spectrometer on Coronas-F, to measure the backscatter
of the flare from the moon without being saturated. Helicon measured
a total backscattered fluence, $f_{bs}=7.5\times 10^{-7} \rm
erg/cm^2$ in the $25-400$ keV range (Golenetskii et al.\ 2004).

The spectrum of the backscattered radiation, a broad hump around
100~keV, is consistent with the spectrum of reflected X-rays and
gamma-rays from cold matter (Lightman \& White 1988). The reflection
of low energy photons below $\sim 50$ keV is suppressed by atomic
absorption (Lightman \& White 1988; George \& Fabian 1991), while
above this energy the dominant process is electron scattering. In
this regime, reflection of photons of increasing energy
($E_\gamma\gtrsim 300$ keV) is progressively more suppressed by the
sharply decreasing Klein-Nishina cross section for backscattering
and by the photon energy loss at each scatter. The albedo for
monochromatic radiation with a normal incidence angle to a semi
infinite plane parallel slab is $\approx 35\%$ for $100$ keV photons
and $\approx 20\%$ for $300$ keV photons, while it drops to a few
percent at $1$ MeV (Imamura \& Epstein 1987). In all these cases the
reflected spectrum peaks below $300$ keV due to the Klein-Nishina
energy loss at each scatter. The sharp drop of the albedo above
$300$ keV implies that unless the spectrum is very hard, the
emerging fluence in the energy window of Helicon closely reflects
the incoming fluence in a similar window.

Assuming a flat spectrum ($F_\nu \propto $ const), we approximate the
backscatter albedo of the moon in Helicon's window as $a \approx 0.25$
(i.e. the ratio between the incoming fluence and the reflected fluence
in the energy window of Helicon). Unless the spectrum is extremely
hard and all the energy is emitted above 0.5 MeV the exact incident
spectral shape and the exact geometry of the scattering should
introduce a correction factor of order unity to this value. Now we can
approximate the fluence of the HGF as:
$S_{25-400}=8S_{bs}/(A_m^2a)\approx 0.3 \;\rm erg/cm^2$ where
$A_m=0.009$ rad is the angular size of the moon from earth.
This fluence implies, using a distance of 15~kpc, an isotropic
equivalent energy of $E_{25-400} \approx 0.8\times 10^{46} \rm erg$.
\footnote{The measurements of fluence and spectrum of the flare was
published (Hurley et al.\ 2005, Palmer et al.\ 2005) while this
paper was in the refereeing process. Our estimate of the fluence in
the $25-400$ keV range is within the errors of these measurements.}

%%%%%%%%%%%%%%%%%%%%%%%%%%%%%%%%%%%%%%%%%%%%%%%%%%

\section{The detection rate of extragalactic hypergiant flares by BATSE}

As young neutron stars, SGRs are remnants of recent core-collapse
SNe. Thus, their abundance should be proportional to the star
formation rate (SFR). Using the SFR in the local universe,
$SFR_{loc}\approx0.02M_\odot ~{\rm y}^{-1}~ {\rm Mpc}^{-3}$ (P{\'
e}rez-Gonz{\' a}lez et al.\ 2003; Schiminovich et al.\  2004), the rate
of HGFs in the local universe can be estimated as:
\begin{equation}\label{EQ HGFRate}
    \dot{N}_{HGF}=2 \times 10^{-4} ~ {\rm y}^{-1} ~{\rm Mpc}^{-3}~
  \frac{\dot{N}_{HGF,MW}}{1/30 {\rm y}^{-1}}\frac{SFR_{loc}}{0.02M_\odot {\rm y}^{-1}
  {\rm Mpc}^{-3}}\frac{3M_\odot{\rm y}^{-1}}{SFR_{MW}},
\end{equation}
where $SFR_{MW}$ is the SFR in the Milky Way ($3_{-2}^{+6} \rm
~M_\odot ~y^{-1}$; Timmes et al.\ 1997; Miller \& Scalo 1979).
BATSE's effective threshold\footnote{
http://cossc.gsfc.nasa.gov/batse/BATSE\_Ctlg/trig\_sen.html} for
triggering on a time scale of $256$ ms is an energy flux of $10^{-7}
\;\rm erg/cm^2/sec$ in the $50-300$ keV energy band\footnote{This
threshold assumes a mildly hard spectrum - a power law with a photon
index of -1.5. The thermal spectrum of the flare, which is
approximately flat in BATSE's window, should introduce a correction
of order unity.} Since the duration of the HGF initial pulse was
shorter than $0.25$ sec, this threshold corresponds to a fluence of
$S_{lim,50-300} \approx 2.5\times 10^{-8} \;\rm erg/cm^2$ .
Fortunately, the albedo of the moon  peaks around BATSE's detection
window, enabling us to estimate the fluence of the HGF at this
window, $S_{50-300}\approx 0.3 \;\rm erg/cm^2$ (\S\ref{sec SGR}).
This implies a maximal distance for detection by BATSE of $\approx
50$ Mpc. Thus, the all-sky rate of similar extragalactic HGFs above
BATSE's detection threshold  is:
\begin{equation}\label{EQ GRBRate}
    \dot{N}^{obs}_{HGF} \approx 130 ~{\rm y^{-1}}
    \left(\frac{R_{SGR}}{15 {\rm kpc}}\right)^3\left(\frac{S_{50-300}/0.3}{S_{lim,50-300}
    /2.5\times10^{-8}}\right)^\frac{3}{2}
  \frac{\dot{N}_{HGF,MW}}{1/30{\rm y}^{-1}}\frac{SFR_{loc}}{0.02M_\odot{\rm y}^{-1}
  {\rm Mpc}^{-3}}\frac{3M_\odot {\rm y}^{-1}}{SFR_{MW}} ,
\end{equation}
where $R_{SGR}$ is the distance to SGR 1806-20. This result is
insensitive to the introduction of beaming (Dar 2005) since in
this case $\dot{N}_{HGF,MW}$ is the rate of HGFs in the Milky Way
which are pointing towards us. This rate is comparable to the
all-sky rate of short GRBs in BATSE's sample,
$\dot{N}^{obs}_{SGRB} \approx 170 \;\rm y^{-1}$, suggesting that
{\it all} short GRBs are extragalactic HGFs.

%%%%%%%%%%%%%%%%%%%%%%%%%%%%%%%%%%%%%%%%%%%%%%%%%%

\section{A search for nearby host galaxies associated with
         short-hard GRBs}

We have compiled a list of all IPN short-hard GRBs with fine
localizations ($3 \sigma$ error box area $\le100$ arcmin$^2$) which
are also at high Galactic latitude ($\delta>20$) to avoid the
complications caused by extinction and source confusion in Galactic
fields. We have found six bursts which satisfy our
criteria\footnote{Note that one of these bursts (GRB 020603) has a
duration of $1.5$ sec in the $25-100$ keV band as observed by
Ulysses (Hurley et al 2002). Since this energy band is similar to
the two lowest BATSE bands and since the duration of GRBs is
typically shorter at higher energy bands, we consider this burst as
a viable short burst. Our results remain unchanged if we exclude
this burst from the sample.}.
%We augment this sample with the first short-hard burst localized by the
%BAT instrument on board the {\it SWIFT} spacecraft, GRB 050202, whose
%error circle has a comparable area.

We first consider the possibility that short-hard GRBs are
physically associated with regions of star formation, as are
long-duration GRBs, soft gamma-ray repeaters (SGRs), and progenitors
proposed by several alternate models. In this framework we obtain a
limit on the distance of a putative host galaxy using the proportionality
of the SFR to the UV luminosity of galaxies in the nearby
universe (Kennicutt 1998). Recent results from the Galaxy Evolution
Explorer (GALEX\footnote{\url http://www.galex.caltech.edu/} show that the
UV luminosity function of nearby galaxies (Wyder et al.\ 2005) is
such that $68\%$[$95\%$] of the UV luminosity (and thus, SFR) is
associated with galaxies brighter than $M_{UV}=-17[-14]$ (which
translates to an SFR of $0.33[0.02] M_{\odot}$ per year). Thus the
hypotheses that a galaxy fainter than $M_{UV}=-17[-14]$ is the host
of a short GRB can be rejected at 1$\sigma$[2$\sigma$]. Given these
limits on the UV brightness of possible hosts, the {\it observed}
luminosities of galaxies within GRB error boxes impose lower limits
on their distances. If nearby, these galaxies would be less luminous than
the limit defined above, and thus unlikely to host a GRB.
The weakest limit is obtained for the brightest galaxy in
the field and we conservatively adopt this as the lower limit on
the distance to each GRB, from which we derive a lower limit on its
energy.

Where possible, we use
available UV imaging from the GALEX mission to find the SFR of
putative host galaxies. In those cases where UV data are not
available, we derive the SFR of each putative host galaxy from its $B$
band magnitude and its $U$-$V$ color (Kennicutt 1998). We verify
that this method produces conservative results when compared to the
GALEX UV analysis. Finally, we briefly consider also the possibility that
short-hard GRBs follow the host galaxies $B$-band or $R$-band
luminosity (rather than the UV/SFR).

GALEX UV data were available for three of the six bursts in our
sample. We have extracted NUV magnitudes and errors for all the
sources detected by GALEX within these error boxes from the
Multy-Mission Science Archive \footnote{\url
http://archive.stsci.edu/} at STScI. We then translated the NUV flux
to SFR (Kenicutt 1998) and deduced the minimal distance to the GRB
by requiring that each galaxy has an SFR above the threshold
described above. In Table 1 we give the minimal distances derived
from the brightest GALEX source in each error box.

For the four localizations for which GALEX data are not available,
we have extracted XDSS POSS II plates \footnotemark. We have also
carried out plate-based analysis for two GRBs for which GALEX data
are available, to check for consistency.
\footnotetext{http://salish.dao.nrc.ca:8080/dss/} Red plates were
available for all fields, and for most of them we have also examined
blue and infrared plates (Table 1). We located the brightest galaxy
within the boundaries of the error box (see Fig. 1 for an example).
In all cases, the brightest galaxy was
the same one in all bands. Additional available data indicated that
several of the nominally brightest galaxies were unlikely to be
nearby host galaxies. In those cases we have considered instead
either the next brightest galaxy in the field or if none was found a
hypothetical galaxy at the plate detection limit. See Table 1 for
details.

Using the XDSS astrometry, we then retrieved the cataloged B and V
magnitudes of the galaxies of interest, using the Naval Observatory
Merged Astrometric Data set (NOMAD; Zacharias et al.\  2004) and
USNO-B1 (Monet et al.\ 2003) catalogs\footnotemark .
\footnotetext{http://www.nofs.navy.mil/nomad/} When available, we
have also extracted $g$, $r$ and $i$ magnitudes from the DPOSS
survey\footnote{http://dposs.ncsa.uiuc.edu} (Djorgovski et al.\
2003) and converted these to $B$ and $V$ magnitudes using
transformations from Smith et al.\ (2002). For two bursts (021201
and 020603), Henden (2002a, 2002b) provides photometric calibrations
of the field. From these, we retrieve accurate $B$ and $V$
magnitudes of the relevant galaxies. The uncertainty in $B-V$ from
Henden is $\delta(B-V)=0.05$ and $\delta(B-V)=0.005$ for the
brightest galaxies in the fields of GRB 021201 and 020603,
respectively. The uncertainty in DPOSS-based $B-V$ values we
estimate as $\delta(B-V)\sim0.1$, including both zero point errors
and the uncertainty introduced by the Smith et al.\ (2002)
transformation. For one burst for which our $B-V$ color is
based solely on NOMAD cataloged values (031214), we set a
conservative error of $\delta(B-V)\sim0.2$, equal to the maximum
offset we find when comparing NOMAD values to either DPOSS or Henden
calibration available for the four other bursts.
%For GRB 050202
%we take an even larger error $\delta(B-V)=0.5$ since the cataloged
%V-band magnitude in NOMAD was inconsistent with the $B$ and $R$
%magnitudes for any galaxy model, and we replaced it with a value
%we derived by averaging the $B-V$ values for galaxies with similar
%$B-R$ colors.
We list our adopted $B$ and $V$ magnitudes and the uncertainty in
$B-V$ for the brightest galaxy detected in each GRB field in Table
1.

Using the correlation derived by Gavazzi, Boselli and Kennicutt
(1991; see their Fig. 1) we use the $B-V$ colors of these galaxies
to derive their $U$-band magnitude and thus their $U-V$ color. With
the $U-V$ color in hand, we use the results of Kennicutt (1998; see
Fig. 2) to derive the star formation rate in each galaxy, in solar
masses per year per $10^{10}$ solar luminosities in the $B$-band.
The $B-V$ uncertainty is properly propagated. Finally, we use this
to find the distance at which the $B$-band luminosity of each
putative host galaxy (derived from the apparent $B$ magnitude and
the distance) is equivalent, through the Kennicutt relation, to an
SFR of $0.33[0.02] M_{\odot}$ per year (our SFR threshold distances appear
in table 2).

To verify the consistency of the SFRs derived from optical colors
using the Kenicutt relation with those derived directly from GALEX
UV data, we conducted both types of analyses for two bursts and
compared the resulting SFR limits. We find that our optical analysis
is conservative, in the sense that the SFR value derived for the
brightest galaxy in each field from optical observations is higher
than the values we get from GALEX UV observations. We are thus
reassured that our analysis of the bursts without GALEX data gives
conservative lower limits on the distance and energy of the relevant
GRBs. As another test of the analysis we have measured the redshift,
$z=0.14$, of the brightest galaxy in the field of GRB\,000607 (Fig.
1). The derived distance, 580 Mpc, is significantly larger than the
minimal distance that we have estimated.

These minimal distance to putative host galaxies result in lower
limits for the isotropic-equivalent energy output of each burst,
$10^{48}-10^{49}$  erg at 2$\sigma$ and $10^{49}-10^{50}$ at
1$\sigma$ (with confidence levels given for each individual burst).
In table 2 we list these distances and the corresponding lower
limits on the bolometric luminosity (when available) and on the
luminosity in the BATSE window for triggering (50--300 keV). The
lower limits we obtain show that none of the six short bursts we
have examined were produced by an event that is similar to the
recent HGF from SGR~1806$-$20.

We have repeated the same analysis (to all the sample except for GRB
790613) assuming that the progenitors of short-hard GRBs follow the
blue light or that they follow the red light.  Namely, we found the
minimal distance to any putative galaxy assuming that its absolute
magnitude is $0.33[0.02]L_*$ in the blue band (using the $B$-band
luminosity function from the 2dF Galaxy Redshift Survey; Croton et
al.\ 2005) and in the red band (using the $R$-band luminosity
function from the Sloan Digital Sky Survey; Blanton et al.\ 2003).
The results in both cases are qualitatively similar to the limits we
obtain under the assumption that the short-hard GRBs follow the SFR.

%%%%%%%%%%%%%%%%%%%%%%%%%%%%%%%%%%%%%%%%%%%%%%%%%%

\section{Discussion}

We have obtained lower bounds on the isotropic equivalent energy
output of six short-hard GRBs under the assumption that their rate
follows the UV luminosity (star formation rate), blue luminosity (as
does the rate of type Ia SNe; Tammann 1970, van den Bergh \& Tammann
1991) or red luminosity of their host galaxies. Under all
assumptions we get comparable limits, suggesting that our results
generally hold in any case that short-hard GRBs are spatially
associated with host galaxies and follow their light \footnote{Note
that this is not the case in the NS-NS merger scenario where bursts
are not necessarily associated with nearby galaxies. In this
scenario the coalescence can occur at a distance of $\sim 200$ kpc
from the host galaxy (Narayan et al.\ 1992). Therefore, for bursts
that are closer than $10$ Mpc, the host galaxies may be outside the
error boxes we examined. However, in the NS-NS merger scenario
bursts are expected to be detectable to much larger distances.}.
With a minimal energy of $10^{48}-10^{49}$ erg in the detection
window of BATSE (Table 1) these bursts are detectable to distances
larger than $\sim 1$ Gpc. Therefore these bursts are part of a
cosmological population of short-hard GRBs, in agreement with the
low $\langle V/V_{max} \rangle$ value of the short-hard GRB
population (Katz \& Canel 1996; Schmidt 2001; Guetta \& Piran 2004).
With a bolometric energy output larger than $ 10^{49}$ erg over a
fraction of a second, their peak luminosities are comparable to
those of long duration GRBs. These results suggest that the
mechanism that produces the majority of short GRBs is related to the
mechanism that produces long GRBs, as is also suggested by the
similarity of the temporal (Nakar \& Piran 2002) and spectral
(Ghirlanda et al.\ 2004) structure of short-hard GRBs and the first
two seconds of long GRBs.

The expected detection rate of extragalactic HGFs similar to the
December 27, 2004 event as short-hard bursts by BATSE (Eq.~\ref{EQ
GRBRate}) is comparable to the total rate of such GRBs. On the other
hand, the lower limits we obtain for our sample of short bursts are
at least two orders of magnitude above the energy emitted during the
December 27 event. Thus, our results suggest that less than $15\%$
of BATSE short-hard GRBs are December 27-type HGFs, and excludes at
$95\%$ confidence level the possibility that more than $40\%$ of the
short GRBs are HGFs.

These findings together leave us with four options, none compelling.
First, it is possible that the short-hard GRB population is in fact
composed of two unrelated phenomena, one local and one cosmological.
By coincidence those two phenomena have comparable detection rates
at the BATSE threshold as well as similar temporal and spectral
properties. In this scenario the local population would
preferentially enrich the sample close to and below the BATSE
threshold. This scenario is only marginally consistent with our null
detection of nearby star-forming galaxies associated with six short
GRBs.

A second possibility is that all short GRBs are
HGFs but that some HGFs emit as much as $10^{49}$ erg of high-energy
radiation. In this scenario the luminosity function of HGFs should
decrease rapidly with energy so that the observed number of HGFs
with $E=10^{46}$ erg would be comparable to that of HGFs with
$E=10^{49}$ erg despite their being visible out to very different
distances. A disturbing question in this scenario is how do isolated
neutron stars produce such gigantic bursts.

A third possibility is that we have observed an unlikely event. That
is, even though the true rate of HGFs is one per SGR lifetime (one
per $\sim 1000$ years), we have seen one after only 30 years of
observations. This implies that HGFs are unique events in the life
of SGRs, as predicted by Eichler (2002), making HGFs a small
fraction of short-hard GRBs.

Finally, a fourth alternative is that the distance to SGR\,1806$-$20 has been
overestimated, and is actually about $6~$kpc. The distance determination of
this SGR is controversial, and based on its association with a star
cluster. Corbel \& Eikenberry (2004) find it to be at $15$ kpc while
Figer et al. (2004) suggest that it is at a distance of $11.8$ kpc,
with the only robust evidence as to its distance being but a lower
limit of $6.4$ kpc (Cameron et al. 2005). At a smaller distance, the
December 27 event is demoted to the rank of bright giant flare. With
energy lower by an order of magnitude compared to the current
estimate, the detection rate of similar extragalactic flares would
decrease by a factor of $\sim$30, making them a small fraction of
short GRBs.

While this paper was being refereed, {\it SWIFT} has localized the
short-hard burst GRB 050509b (Geherles et al. 2005). While the
identity of the host galaxy of this burst has not been conclusively
determined (e.g., Bloom et al 2005, Fox et al 2005)
it is clearly at or above $z=0.22$ (D=900 Mpc).
Thus, this event strengthens our results, suggesting that most
short-hard bursts are at cosmological distances. Furthermore, the
low localization rate of short-hard bursts achieved by {\it SWIFT}
so far ($\lesssim 4~{\rm yr}^{-1}$) suggests that the sample size of
well-localized short-hard bursts is not likely to rapidly increase.
Statistical studies of host galaxies will therefore probably
continue to rely heavily on ``historical'' short-hard burst samples,
such as the one we have analyzed here.

%\section*{Acknowledgements}

We would like to thank E. S. Phinney, R. Sari, M. Kamionkowski,  S.
Furlanetto, J. Katz, A. Pickles and S. Eikenberry, for helpful
discussions and remarks. We thank J.L Atteia for directing our
attention to GRB 790613. A.G. acknowledges support by NASA through
Hubble Fellowship grant \#HST-HF-01158.01-A awarded by STScI, which
is operated by AURA, Inc., for NASA, under contract NAS 5-26555.
This research was partially supported by a grant from the US-Israel
BSF.

%%%%%%%%%%%%%%%%%%%%%%%%%%%%%%%%%%%%%%%%%%%%%%%%%%

\clearpage

%%%%%%%%%%%%%%%%%%%%%%%%%%%%%%%%%%%%%%%%%%%%%%%%%%

\newcommand{\tb}[1]{{\small {#1}}}

\begin{table*}
\caption{The brightest galaxies within short-hard GRB error boxes}
\label{tbl1}
 \begin{tabular}{cccccccc}%lclclclclclclclc}

 \hline
    GRB & XDSS   & RA      & Dec & NUV   & B     &  B-V   & U-V     \\
        & Plates & (J2000) &             & [mag] & [mag] &  [mag] & [mag]   \\
 \hline

031214$^{a}$ & R,I    & 16:21:44.38 & -11:43:07.3 & - &$18.09 {\tb \pm 0.3}$ & - & -  \\
021201$^{b}$ & B,R,I  & 08:07:44.33 & 21:14:30.1 & - & $18.88{\tb \pm 0.05}$ & $1.52{\pm 0.05}$ &$  3.1{\pm 0.11}$ \\
020603$^c$ & R,I& 15:46:29.36 & -22:17:09.9 & -  & $16.03{\pm 0.006}$& $1.056{\pm 0.005}$& $1.68{\pm 0.01}$ \\
 &  &15:46:34.50  &-22:14:52.9 & - & $16.524{\pm 0.002}$ & $0.625{\pm 0.004}$ & $0.71{\pm 0.01}$ \\
001204 & B,R,I  & 02:41:04.36 & 12:52:39.2  &  $21.3{\pm 0.2}$ &$19.87\pm 0.07$ & $0.46\pm 0.1$ & $0.34 \pm 0.2$       \\
000607$^d$ & B,R,I  & 02:33:48.50 & 17:03:59.5  & $22.2{\pm 0.5}$ & $18.15\pm 0.07$ & $0.4\pm 0.1$ & $0.2 \pm 0.2$        \\
   &  & 02:33:47.62 & 17:03:28.9 & $22.1{\pm 0.4}$ &  - & - & -         \\
790613 & - & 14:11:59.93 & 78:41:07.0 & $21.4{\pm 0.2}$ &  - & - & -         \\
%050202 & $ R $ & 19:21:57.88 & -38:43:52.2 & - & $15.9{\tb \pm 0.2}$  & $1.27{\tb \pm 0.5}$  &  $xx.xx{\tb \pm 0.x}$         \\

 \hline

\end{tabular}
\medskip

{\small a) This is the magnitude of the brightest clearly resolved
galaxy. However, on this plate it was difficult to distinguish
nearly-saturated stars from compact round galaxies. We therefore
conservatively adopt the brightest nearly-saturated/compact object
as our ``brightest galaxy'' ($B\sim17$) and assign it (again,
conservatively) a very blue color ($U-V=0$ mag).

b) Both the brightest galaxy and the only other detectable galaxy
within the error box, have very red cataloged colors (yielding
values of $U-V>2.5$), and are thus expected to have negligible SFR.
We have therefore conservatively assumed that the putative host is
blue ($U-V=0$) and it lies just below the blue plate limit ($B\sim
20.5$).

c)The brightest galaxy is rather red ($U-V \approx 1.7$ mag) and
therefore it is not the galaxy with the largest SFR in the field.
The properties of the brightest blue galaxy in the field are listed
as well.

d)We have observed the only optically detectable galaxy in this
error box using the double spectrograph mounted on the 200''
telescope at Palomar Observatory. We determine the redshift of this
galaxy to be $z=0.14$, based on detected emission lines (H$_\alpha$
and NII). The lower bound on the distance of a putative host galaxy
is derived from a second galaxy with comparable NUV brightness.

}

%\end{minipage}
\end{table*}

\clearpage

\begin{table*}
%\begin{minipage}{175mm}
 \caption{The minimal distances to putative star-forming host galaxies
          and corresponding energetics}
  \label{tbl2}
 \begin{tabular}{ccccccccc}%lclclclclclc}
 \hline
    GRB & $S^a_{0.015-5\rm{MeV}}$  & $S^a_{50-300\rm{keV}}$ & $D^b_{min}$ & $E^c_{min,0.015-5\rm{MeV}}$  & $E^c_{min,50-300\rm{keV}}$ & Ref. \\
 \hline
 031214 & $3 \times 10^{-4}$ & $4 \times 10^{-5 \dagger}$ & $80~~ [20]$ & $20~ [1.4]$ & $2~~ [0.14]$ & {\small Hurley et al.\ 2003}\\

021201 & - & $6 \times 10^{-7\dagger\dagger}$ & $400~ [100]$ & - &
$1~ [0.06]$ & {\small Hurley et al.\ 2002d}\\ %9 arcmin^2

020603 & - & $2 \times 10^{-5\dagger\dagger}$ & $100 ~~[25]$ & - &
$2~~[0.14]$ & {\small Hurley et al.\ 2002b,c}\\ %19 arcmin^2

001204 & $2 \times 10^{-6}$ & $5 \times 10^{-7}$ & $400 ~[100]$ &
$4~ [0.2]$& $1 ~[0.06]$ & {\small Hurley et al.\ 2002a}\\ %6 arcmin^2

 000607 & $5 \times 10^{-6}$ & $5 \times 10^{-7}$ & $480~ [120]$ & $12~[0.8]$& $1.2~[0.08]$ & {\small Hurley et al.\
 2002a}\\ %5.6 arcmin^2

790613 & $3 \times 10^{-6}$ & $-$ & $400~[100]$ & $6~ [0.4]$& -
  & {\small Barat et al.\ 1984}\\ %0.7 arcmin^2

%050202 & $x \times 10^{-x}$ & $-$ & $x$ & $x \times 10^{x}$ & -
%  & {\small Sakamoto et al.\ 2005}\\ \hline

\end{tabular}
\medskip

{\small a) The fluence in $\rm erg/cm^2$

b) The minimal distance to the burst in Mpc, at $1 [2]\sigma$
confidence level, assuming that the rate of short-hard GRBs follow
the SFR.

c) The minimal isotropic energy output of the bursts in units of
$10^{49}$ erg at $1 [2]\sigma$ confidence level.

$^\dagger$) Calculated from the fluence in 15-5000keV assuming a low
frequency photon index $\alpha=-1$ and taking a peak frequency
$E_p=2000$ (Hurley et al.\ 2003).

$^{\dagger \dagger}$) Calculated from the fluence in 25-100 keV
assuming a low frequency photon index $\alpha=-1$ and a peak
frequency $E_p>300$

 }
%\end{minipage}
\end{table*}

\clearpage

\begin{figure}[h]
\begin{center}
\includegraphics[width=14cm,height=14cm]{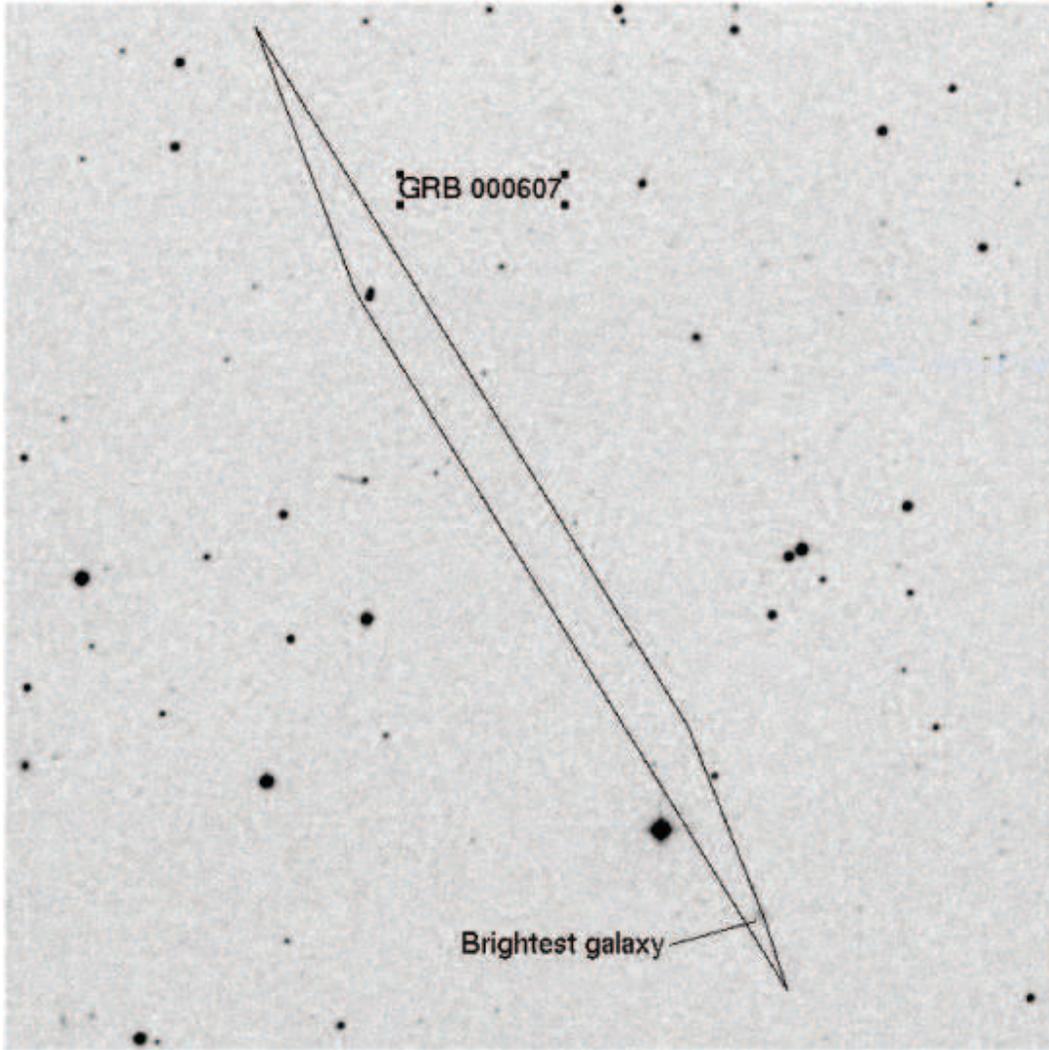}
\caption{The error box of GRB 000607 superposed on a blue POSS II
plate. The brightest galaxy turns out to be at z=0.14. All other
sources within the error box have spatial light distributions
consistent with those of stars.} \label{plotone}
\end{center}
\end{figure}

%%%%%%%%%%%%%%%%%%%%%%%%%%%%%%%%%%%%%%%%%%%%%%%%%%

\end{document}